\documentclass[a4paper,UKenglish,cleveref, autoref, thm-restate.]{lipics-v2021}

\hideLIPIcs  %

\usepackage{amssymb}
\usepackage{times}
\usepackage{amsmath}
\usepackage{makeidx}  %
\usepackage{graphicx,color}
\usepackage{floatflt,wrapfig}
\usepackage{listings}

\usepackage{doi}
\usepackage[export]{adjustbox}
\usepackage{here}

\usepackage{diagbox}
\usepackage{extarrows}
\usepackage{stmaryrd}
\usepackage{tabularx}
\usepackage{cleveref,todonotes}

\usepackage{paralist}
\usepackage{color}
\usepackage{alltt}
\usepackage{moreverb}
\usepackage{multirow}
\usepackage{algorithmic}
\usepackage{listings}
\usepackage{setspace}
\usepackage{wrapfig,lipsum,booktabs}
\usepackage{colortbl}
  \usepackage{eurosym}
  \usepackage{framed}
\usepackage{pdfpages}
\usepackage{xspace}
\usepackage{bussproofs}
\usepackage{hyperref}
\usepackage{color,soul}

\usepackage{tikz}
\usetikzlibrary{arrows}
\usetikzlibrary{positioning}

\usepackage[utf8]{inputenc}
\usepackage[T1]{fontenc}

\usepackage{color}
\definecolor{keywordcolor}{rgb}{0.7, 0.1, 0.1}   %
\definecolor{tacticcolor}{rgb}{0.0, 0.1, 0.6}    %
\definecolor{commentcolor}{rgb}{0.4, 0.4, 0.4}   %
\definecolor{symbolcolor}{rgb}{0.0, 0.1, 0.6}    %
\definecolor{sortcolor}{rgb}{0.1, 0.5, 0.1}      %
\definecolor{attributecolor}{rgb}{0.7, 0.1, 0.1} %

\lstdefinestyle{leanstyle}{
  language=lean,
  numbers=left,
  numbersep=10pt,
  tabsize=2,
  showspaces=false,
  showstringspaces=false
}

\usepackage{url}

\newcommand\solver[1]{\textsc{#1}\xspace}

\newcommand\vampire{\solver{Vampire}}

\newcommand\lean{\solver{Lean}}
\newcommand\avatar{\solver{Avatar}}
\title{Lean on Vampire Proofs (Short Paper)} %

\author{Jonas Bodingbauer}{TU Wien, Austria }{jonas.bodingbauer@tuwien.ac.at}{https://orcid.org/0009-0006-2607-4325}{}
\author{M\'arton Hajdu}{TU Wien, Austria}{marton.hajdu@tuwien.ac.at}{https://orcid.org/0000-0002-8273-2613}{}
\author{Laura Kov\'acs}{TU Wien, Austria}{laura.kovacs@tuwien.ac.at}{https://orcid.org/0000-0002-8299-2714}{}
\author{Axel Polaczek}{TU Wien, Austria}{axel.polaczek@tuwien.ac.at}{https://orcid.org/0000-0002-7439-5987}{}
\author{Michael Rawson}{University of Southampton, UK}{michael@rawsons.uk}{https://orcid.org/0000-0001-7834-1567}{}

\authorrunning{Bodingbauer, Hajdu, Kov\'acs, Polaczek, and Rawson} %

\Copyright{} %



\category{} %

\relatedversion{} %

\acknowledgements{We thank Henrik Böving for providing helpful insights on \lean{}. 
This research was funded in whole or in part by the  ERC Consolidator Grant ARTIST 101002685, the ERC Proof of Concept Grant LEARN 101213411, and by SBA Research (SBA-K1 NGC), a COMET Center within the COMET – Competence Centers for Excellent Technologies Programme, and funded by BMIMI, BMWET, and the federal state of Vienna. The COMET Programme is managed by FFG
.}%

\nolinenumbers %

\EventEditors{John Q. Open and Joan R. Access}
\EventNoEds{2}
\EventLongTitle{42nd Conference on Very Important Topics (CVIT 2016)}
\EventShortTitle{CVIT 2016}
\EventAcronym{CVIT}
\EventYear{2016}
\EventDate{December 24--27, 2016}
\EventLocation{Little Whinging, United Kingdom}
\EventLogo{}
\SeriesVolume{42}
\ArticleNo{23}

\begin{document}

\maketitle

\begin{abstract}
\vampire proves theorems completely automatically in first- and higher-order logic extended with theories. Proof checking is increasingly demanded to consolidate user trust in \vampire's output.  We describe ongoing efforts in reconstructing \vampire proofs as trusted proofs in \lean.
\end{abstract}

\section{Introduction}
Computer-assisted proofs are indispensable in applications of mathematics~\cite{DBLP:journals/jar/McCune97,DBLP:conf/aisc/UrbanS10,bolan2025equationaltheoriesprojectadvancing}, cybersecurity~\cite{DBLP:conf/cav/Rungta22,DBLP:conf/sp/JeanteurKMR24,DBLP:conf/ccs/LauferOB24}, 
software analysis~\cite{DBLP:conf/atva/BeutnerF24,DBLP:conf/jelia/AhrendtBBGHHMS00,DBLP:conf/fmcad/GeorgiouGBRKR22},  neural networks~\cite{DBLP:journals/cacm/BarrettHS26,DBLP:conf/itp/DesmartinIPKSK25}, and many other areas. Automated reasoners, such as automated theorem provers (ATPs) and satisfiability modulo theory (SMT) solvers, provide push-button solutions to machine-based proof generation. Proofs returned by automated reasoners are can be checked by e.g. using an external ATP/SMT solver to reprove each step of the generated proof~\cite{DBLP:conf/cade/RawsonVSK25,DBLP:conf/tacas/LachnittFARBNBT24,10.1007/978-3-642-25379-9_12}. Alternatively, one can rely on interactive theorem proving (ITP) to replay the proof inference-by-inference~\cite{DBLP:journals/jar/MengP08,DBLP:conf/cav/MohamedMKBRQTB25}.
Unfortunately, proof-checking efforts struggle to accommodate advanced reasoning features: for example, quantifier elimination cannot be checked by SMT solvers, inductive proofs are often not supported by existing proof standards, and equational reasoning modulo associativity and commutativity (AC) is very difficult to recover in ITPs. There is a constant tension between advances in ATP~\cite{CVC52022,spass,DBLP:conf/fm/Passmore21,DBLP:conf/cade/0001CV19,DBLP:conf/cade/DuarteK20,DBLP:conf/ijcar/KovacsHHV24} and proof checking. %

\noindent{\bf Our contributions.} This paper addresses proof checking for the \vampire theorem prover~\cite{DBLP:conf/cav/BartekBCHHHKRRRSSV25}, making proof checking feasible and adjustable. We enhance \vampire  with \emph{trusted proofs}:  proofs generated by \vampire{} are reconstructed in, and checked by, the %
\lean interactive theorem prover~\cite{DBLP:conf/cade/Moura021}. 
We 
describe our ongoing efforts in extending \vampire{} with \lean-based proof formats (\Cref{sec:method}), allowing us to use \lean for increasing user trust in \vampire{} proofs. Our experiments showcase scalability of using \lean for trusting \vampire (\Cref{sec:experiments}).

\noindent {\bf Related methods} have emerged from both ATP and ITP communities. We complement the interfacing efforts of Lean-auto~\cite{DBLP:conf/cav/QianCBA25} by providing tactics for validating \vampire proofs in \lean.  Compared  to~\cite{DBLP:conf/cade/RawsonVSK25}, trust in \vampire proofs is handed off to \lean, rather than an SMT solver, significantly reducing the trusted computing base.
Work with SMT solvers~\cite{DBLP:conf/cav/MohamedMKBRQTB25} is similar, but here we consider superposition-based theorem proving.
This is the first step towards a long-term vision turning \vampire %
into a hammer for \lean, complementing existing efforts in ATP/ITP integration~\cite{zhu2025premiseselectionleanhammer,VampireDedukti} and continuing a long tradition of \vampire service in hammer systems~\cite{DBLP:conf/itp/DesharnaisVBW22}.

\section{Illustrative Example and %
Scope}\label{sec:motiv}
\newcommand\mul{\mathsf{mul}}
{\small
\begin{figure}[t]
  \small
  \newcommand\ProofAnnot[1]{\text{(#1)}}
  \newcommand\ProofStep[3]{#1\hspace*{-.75em} & #2 &\hspace*{-.85em} \text{#3}\\}
  
  \newcommand\ident{\mathsf{id}}
  $$
  \begin{array}{l||l}
  \hspace*{-1em}
  \begin{array}{lll}
    \ProofStep{1.}{\mul(x,\mul(y,z))=}{Input}
    \ProofStep{}{\qquad\qquad\mul(\mul(x,y),z)}{}
    \ProofStep{2.}{ \mul(\ident,x)=x}{Input}
    \ProofStep{4.}{ \ident=\mul(x,x)}{Input}
    
    \ProofStep{6.}{\lnot \forall{x,y} : \mul(x,y) = \mul(y,x)}{Neg. Goal}
    \ProofStep{7.}{\exists{x,y} : \mul(x,y) \neq \mul(y,x)}{ENNF 6}
  \ProofStep{8.}{{ \mul(\sigma_0,\sigma_1)\neq\mul(\sigma_1,\sigma_0)}}{{Skolem 7}}
  \ProofStep{15.}{\mul(x,\mul(x,y)) = \mul(\ident,y)}{Sup~1,4}
    
     \end{array}
  &
  \begin{array}{lll}
    \ProofStep{19.}{\ident = \mul(x,\mul(y,\mul(x,y)))}{Sup~4,1}
    \ProofStep{22.}{\mul(x,\mul(x,y))=y}{FwDem~15,2}
    \ProofStep{25.}{\mul(x,\ident)=x}{Sup~22,4}
    \ProofStep{80.}{\mul(y,\mul(x,y))=\mul(x,\ident)}{Sup~22,19}
    \ProofStep{85.}{\mul(y,\mul(x,y))=x}{FwDem~80,25}
    \ProofStep{160.}{ \mul(x,y)=\mul(y,x)}{Sup~22,85}
    \ProofStep{249.}{\mul(\sigma_0,\sigma_1) \neq \mul(\sigma_0,\sigma_1)}{Sup~8,160}
    \ProofStep{250.}{\text{{\bf false}}}{Resolve 249}
  \end{array}
  \end{array}
  $$
  \vspace{-1.8em}
  \caption{Simplified \vampire proof showing that if every element of a group has order two, the group is commutative. Here,  $\mul$ denotes the group operator and $\ident$ the group's identity element. The variables $x,y$ are implicitly universally quantified, whereas $\sigma_0$ and $\sigma_1$ are Skolem constants. The proof steps $\text{Sup, FwDem}$  respectively denote superposition and its variant forward demodulation.\vspace*{-1em}}%
  \label{fig:group-proof}
\end{figure}}
\noindent{\bf Saturation and \vampire.} Proof automation with {\vampire} is performed via saturation~\cite{DBLP:conf/cav/KovacsV13}: given a conjecture $F$ in (extensions of) first-order logic, \vampire{} negates and clausifies $\lnot F$ and tries   
to derive falsum from the resulting clause set by saturating the set with respect to the the superposition calculus~\cite{DBLP:books/el/RV01/BachmairG01,DBLP:conf/cade/NieuwenhuisHRV01}. During proof search, clauses may be removed if they are determined to be \emph{redundant}. Whenever the empty clause ${\bf false}$ is derived as a consequence of $\neg F$, (classical) validity of $F$ is established and a proof is generated as part of the \vampire{} output.
\Cref{fig:group-proof} shows a sample proof in which \vampire applies specializations of superposition inferences, such as forward demodulation.\smallskip

\noindent{\bf Checking \vampire proofs for soundness.} 
To improve efficiency of proof search \vampire implements dedicated inference rules for equality reasoning, redundancy checking, arithmetic, and induction~\cite{DBLP:conf/cav/BartekBCHHHKRRRSSV25}, among others. These advances increase the number of inference rules in \vampire to around 200. %
Our work   identifies a core set of inference rules (\Cref{sec:method}) from which trusted proofs of \vampire are generated. Trust in \vampire proofs is achieved by  the help of \lean: %
 (i) we extend \vampire to output proofs as \lean theorems proven using lemmas corresponding to inferences. Once \vampire finds a proof, a \lean input file is generated where tactics are  used to (ii)  {\it reconstruct  \vampire inferences in \lean}, demonstrating soundness for each step. A final combination of these steps results in a full \vampire proof from assumptions to conclusions trusted by \lean.
For example, each \vampire rule from \Cref{fig:group-proof} is converted to a \lean theorem and proven by a collection of dedicated tactics, such as the superposition tactic in \Cref{listing:superposition_theorem}. %

\section{\lean Trust on  \vampire Proofs}\label{sec:method}

We produce \lean-checkable proofs for  problems solved by \vampire (\Cref{sec:imple}).
We report on our efforts in reconstructing the preprocessing steps (\Cref{sec:preprocess}) and the most important inference rules (\Cref{sec:infRules}) of \vampire in \lean. While this paper focuses on standard  first-order logic with equality, extending our work to multi-sorted first-order logic is an immediate step for future work. 

\begin{subsection}{Implementation}\label{sec:scope}\label{sec:imple}

Our approach for generating trusted \vampire proofs is implemented\footnote{Our implementation is open-source and available on GitHub: \href{https://github.com/vprover/vampire/tree/leancheck}{https://github.com/vprover/vampire/tree/leancheck}, Lean library \href{https://github.com/vprover/vamplean}{https://github.com/vprover/vamplean}} as a shared effort between \vampire and \lean: once a \vampire proof is generated, we output a \lean file in which we construct proof terms in \lean using tactics provided by \lean and mathlib\cite{Mathlib}. 
Generating proof terms can  be  assisted by  additional information gained through proof replay in \vampire, for example in the case of unification or matching. For the frequently used inference rules  in saturation (\Cref{sec:infRules}), clauses are instantiated via the most general unifiers (mgu) computed by \vampire and  passed to the \texttt{grind} tactic in \lean. \smallskip

\noindent{\bf{General structure of trusted \vampire proofs.}} 
A  \lean file representing a trusted \vampire proof is structured in three parts. After the preamble that declares all required symbols, the second (typically largest) section includes all individual inference steps as theorems. Each theorem is proven using a collection of \lean tactics, or in very simple cases directly by providing suitable proof terms. Tactics are often needed, because \vampire inferences do not keep the ordering of clauses and hence require suitable AC rule applications. 
The third part is a proof of the input conjecture, which shows that the set of axioms imply the conclusion. This is done by using the previously proven lemmas to reconstruct the proof found by \vampire and also includes symbol introduction, such as Skolemization and Tseitin transformation, which are part of \vampire's preprocessing (see \Cref{fig:group-proof}). 
Formulas are encoded in a natural way in \lean: the primitives of first-order formulas are used as defined by \lean core and mathlib \cite{Mathlib}. Since \vampire treats all domains as inhabited, this must be taken into account in \lean, which allows empty sorts.
{ 
\begin{lstlisting}[language=lean,
caption={Third part of a \lean input file, representing a trusted \vampire proof corresponding to \Cref{fig:group-proof}. It is combining the previously proven lemmas corresponding to inference steps for the full proof of \Cref{fig:group-proof}.},
label=listing:second_part_proof,
basicstyle=\fontsize{8}{9}\ttfamily,
]
-- inference steps above
theorem fullProof : 
(∀ v0 v1 v2 : ι, ((mul (mul v0 v1) v2)=(mul v0 (mul v1 v2)))) → 
(∀ v0 : ι, ((mul id v0)=v0)) → (∀ v0 : ι, (id=(mul v0 v0))) → 
(∀ v0 v1 : ι, ((mul v0 v1)=(mul v1 v0))) := by
  intros step1 step2 step3 
  apply Classical.byContradiction; intro step5
  ...
  have step160 := inf_s160 step22 step85
  have step249 := inf_s249 step8 step160
\end{lstlisting}}
\end{subsection}

\begin{subsection}{Preprocessing}\label{sec:preprocess}
Several preprocessing steps are used in \vampire. To transform a first-order formula $F$ (i.e.\ conjecture) into clausal normal form (CNF), the formula $F$ is (typically) rectified, simplified, flattened, transformed to equivalence negation normal form (ENNF),\footnote{Equivalence NNF allows $\leftrightarrow$ and $\oplus$, with negations in front of atoms.} named, converted to NNF, Skolemized, and finally transformed to conjunctive normal form~\cite{GCAI2016:New_Techniques_Clausal_Form}. For ENNF transformation as well as flattening and simplification, we use the \lean \texttt{simp} tactic, with suitable theorems.\footnote{In (E)NNF transformation, \vampire by default uses non-confluent rewrite rules, however its outermost reduction strategy can be replicated using the $\downarrow$ annotation.} For the transformation from ENNF to CNF, our implementation currently uses either \texttt{duper}~\cite{clune_et_al:LIPIcs.ITP.2024.10} or \texttt{grind}. \smallskip %

\noindent{\bf{Skolemization.}}
After the input conjecture $F$ has been transformed to NNF, the existential quantifiers are replaced with Skolem functions of suitable arity. The default implementation of \vampire replaces existentially-quantified variables by functions that are dependent on free variables of the immediate subformulas. Automating this behavior in proof reconstruction is challenging; we therefore implemented an option to change \vampire Skolemization to use all currently bound variables (which is always a superset of the previously bound variables). For showing correctness of the employed Skolemization mechanism, we prenex the existential quantifiers using the theorem \texttt{Classical.skolem} of type: 

$$(\forall (x : \alpha), \exists (y : b\ x), p\ x\ y) \leftrightarrow \exists (f : (x : \alpha \rightarrow b\ x), \forall (x : \alpha), p\ x\ (f\ x)).$$
Additional theorems that allow prenexing an existential from both $\vee$ and $\wedge$ are also used. For the formula in existential prenex form, each of the quantifiers can then be assigned a new function name. Special care needs to be taken to match the ordering of Skolem functions assigned to variables as well as the orientation of equalities which is handled by custom tactics.
The Skolemization step~8 of the running example shown in \Cref{fig:group-proof} is given  in~\Cref{listing:skolemization}.
\begin{lstlisting}[language=lean,
caption={Skolemization of the two existentially quantified variables in step~8. This snippet is from the third part of a trusted proof and  assembles all inferences. \texttt{symm\_match} and \texttt{exists\_prenex} are custom tactics which take care of the orientation of equalities and prenexing.},
label=listing:skolemization,
basicstyle=\fontsize{8}{9}\ttfamily,
]
  -- step8 skolemisation
  exists_prenex at step7
  let ⟨sK0,sK1,step8'⟩ := step7
  have step8 : (¬((mult sK0 sK1)=(mult sK1 sK0))) := by symm_match using step8'
\end{lstlisting}

\noindent{\bf{Predicate and function definitions.}}
\vampire may introduce predicates naming subformulas. These are declared globally as implicit variables, later defined in the final part of the \lean input. Additionally, a suitable equivalence theorem is added and used by later inference steps from \vampire.
\end{subsection}

\begin{subsection}{Inference Rules within Saturation}\label{sec:infRules}
Saturation-based proof search in \vampire implements the superposition calculus and uses the following core rules: resolution, factoring, paramodulation, equality resolution, and equality factoring. Many rules applied in \vampire, such as  superposition and (forward) demodulation, can be described as a restricted instance of paramodulation. %

\Cref{listing:superposition_theorem} demonstrates superposition used by \vampire at step~160 of \Cref{fig:group-proof}. It is translated as a \lean theorem that takes two premises and  gives a conclusion. At the beginning of the theorem, all premises are introduced as well as the variables of the conclusion. The most general unifier (mgu) found by \vampire is instantiated via the \texttt{have} statements. After that, the \texttt{grind} tactic is called, which is well-suited for ground reasoning and  can effectively reconstruct  proof terms.\footnote{This approach is inspired by~\cite{DBLP:conf/cade/RawsonVSK25}, where ground instances are passed to an SMT solver.}

Other inference rules of \vampire follow the same steps: introduce premises and conclusion variables, instantiate them according to the unifier found by \vampire,  and then dispatch it to \texttt{grind}. When it occurs that two variables are unified, the unifier can contain a variable that is not present in the conclusion~\cite{DBLP:conf/cade/RawsonVSK25}. In this case, the default element of the inhabited sort is used. 
In the future, more specialized, custom, tactics could be employed that can use additional information provided by \vampire such as the positions where rewrites occurred.\footnote{\vampire does not preserve literal ordering, therefore a tactic that takes into account AC of $\lor$ is required.}
\begin{figure}[H]
\begin{minipage}{0.54\textwidth}
\begin{lstlisting}[language=lean,
basicstyle=\fontsize{8}{9}\ttfamily,
]
theorem inf_s160 : 
 (∀ x₀ y₀ : ι, ((mul y₀ (mul x₀ y₀))=x₀)) →
 (∀ x₁ y₁ : ι, ((mul x₁ (mul x₁ y₁))=y₁)) →
 (∀ x  y  : ι, ((mul x y)=(mul y x))) := by
 intros h0 h1 x y
 have i0 := h0 x y
 have i1 := h1 y (mul x y)
 grind only [cases Or]
\end{lstlisting}
\begin{center}(a) \end{center}
\end{minipage}
\hspace{1em}
\begin{minipage}{0.44\textwidth}
{\small
\qquad General paramodulation rule\\[.2em]
\AxiomC{$l = r \lor C$}
\AxiomC{$L[k] \lor D$}
\BinaryInfC{$(L[r] \lor C \lor D)\sigma$}
\DisplayProof
$\sigma = \operatorname{mgu}(l, k)$
}
\\[0.5em]
\qquad{\small \text{instantiated with}}\\[0.5em]
{
\footnotesize
\setlength{\tabcolsep}{5pt}
\begin{tabular}{c|l}
     $l$ & $\mul(y_0,\mul(x_0,y_0))$ \\
     $r$ & $x_0$\\
     $L[k]$ & $\mul(x_1,\mul(x_1,y_1)) = y_1$\\
     $\sigma$ & $\{x_0 \mapsto x, y_0 \mapsto y,$\\
     &          $\ \ x_1 \mapsto y, y_1 \mapsto \mul(x,y)\}$
\end{tabular}
}
\begin{center} (b) \end{center}
\end{minipage}
\captionsetup{type=lstlisting}
\caption{
(a) \lean theorem representing the superposition step~160 of \Cref{fig:group-proof}. 
(b) The  instantiation of  paramodulation (i.e.\ more general superposition)  for (a). %
Clauses $C, D$ are empty for the instance of \Cref{fig:group-proof}.}
\label{listing:superposition_theorem}
\end{figure}
\end{subsection}

\begin{subsection}{AVATAR}
A feature that contributes to \vampire's success is 
\avatar~\cite{Voronkov:AVATAR:CAV:2014}, which enables efficient reasoning over the propositional/quantifier-free structure of clauses. %
Simply put, \avatar  splits clauses into variable-disjoint sub-clauses and defines propositional labels for each of the sub-clauses, as follows. \smallskip

\noindent{\bf{\avatar splitting of a clause:}}
{
\small
\AxiomC{$\neg Q(z) \lor P(x, f(z))\lor R(y)$}
\UnaryInfC{$\mathrm{sp}_1 \lor \mathrm{sp}_2$}
\DisplayProof
{\normalsize
\ using
}
\begin{tabular}{l}
\(\mathrm{sp}_1 \equiv \neg Q(z) \lor P(x, f(z))\)\\
\(\mathrm{sp}_2 \equiv R(y)\)
\end{tabular}
} 

\noindent A SAT solver in \avatar is used to reason about the propositional structure $\mathrm{sp}_1\lor \mathrm{sp}_2$,  while \vampire handles  \avatar components of the form \(R(x) \leftarrow \mathrm{sp}_2\),  which should be understood as ``$R(x)$ if $\mathrm{sp}_2$''. 
If the SAT solver finds that the propositional structure is unsatisfiable, a refutation  is produced. When the SAT solver finds a model, \vampire  continues reasoning using the learned knowledge from the propositional model and derives new contradictions using first-order reasoning, that can in turn be used to define new propositional clauses to be passed to the SAT solver. %
When a \vampire proof is found with \avatar, the final contradiction (${\bf false}$) is derived by the SAT solver.\smallskip

\noindent{\bf \avatar in \lean.} To reconstruct and check \avatar steps in \lean, the propositional names are declared and defined in the first and last parts of \vampire proofs (\Cref{sec:imple}).  When a proof step involves an \avatar component, the propositional labels of the conclusions are introduced before introducing the propositional variables (e.g. $\mathrm{sp}_1$); these labels are used to instantiate  premises, as shown in \Cref{listing:superposition_theorem_avatar}. The other rules are treated as normal inference rules, where the splitting clauses is treated using rewrites as well as \texttt{grind}, which only needs to handle AC reordering as well as removing potentially duplicate literals. The final SAT refutation is handled using the \texttt{bv\_decide} tactic, which calls a SAT solver and then verifies the received proof within \lean \cite{bvdecide_tactic}.

\begin{lstlisting}[language=lean,
basicstyle=\fontsize{8}{9}\ttfamily,
caption={Variant of the inference rule of \Cref{listing:superposition_theorem} if it were part of an \avatar proof. The final SAT proof, which is reconstructed using \texttt{bv\_decide}, is also listed.},
label=listing:superposition_theorem_avatar
]
theorem inf_s160_avatar : (sP1→(∀ x₀ y₀ : ι, ((mul y₀ (mul x₀ y₀))=x₀))) →
 (∀ x₁ y₁ : ι, ((mul x₁ (mul x₁ y₁))=y₁)) → 
 (sP1→(∀ x  y  : ι, ((mul x y)=(mul y x)))) := by
 intros h0 h1 sp1 x y
 have i0 := h0 sp1 x y
 ...
theorem avatarRefutation' (sP1' sP2' sP3' : Bool) : (sP1' || sP2' || sP3') → (!sP2') → (!sP3') → (!sP1') → False := by bv_decide
\end{lstlisting}
\end{subsection}

\pagebreak
\begin{section}{Experiments}\label{sec:experiments}

\begin{wrapfigure}{r}{0.5\textwidth}
\footnotesize
\begin{tabular}{l|c|ccc}
\multirow{2}{*}{TPTP} & \multirow{2}{*}{\vampire} & \multicolumn{3}{c}{\lean} \\
\cline{3-5}
 &                           & Success & Timeout & Error \\
\hline
CNF      &            3860           &    3785 &   53    &   22        \\
FOF      &            3897           &    3296 &   482   &   119       \\
\end{tabular}
\caption{Number of proofs found by baseline \vampire (column~2) vs our  \lean proof reconstructions (columns~3--5),  sorted for success/failure reasons. Success means trusted \vampire proofs, validated by \lean.\label{table:exp}}
\end{wrapfigure}

Our experiments used benchmarks from version 9.2.1 of the TPTP library~\cite{TPTP}. We relied on  \textsc{BenchExec}~\cite{benchexec} for conducting  experiments, using an AMD Epyc 7502 2.5GHz processors and 1TB RAM. Each benchmark ran with a single core and 16GB of memory. Our benchmark set consists of all CNF and FOF problems from TPTP, overall 17,603 problem instances. Our \vampire{} runs used \vampire{} version 5.0.1 with the Discount saturation loop and a 20-giga-instruction (roughly 5-10 seconds) resource limit (\texttt{-sa discount -i 20000}); we refer to this \vampire version as baseline \vampire. For \lean{}, we used a 300-second timeout.

We compare  (i) baseline \vampire{} without  proof output ({\tt --proof off}) and (ii)  our work (\Cref{sec:imple}) enhancing   \vampire (Git tag: \href{https://github.com/vprover/vampire/releases/tag/Leancheck2026}{Leancheck2026}) with output of proofs checkable by \lean. \Cref{table:exp} summarizes our results, showcasing  98\% success rate of trusted proof generation on CNF problems and 85\% of FOF examples. We note that CNF proof reconstruction reduces the load of reconstructing preprocessing steps. For problems whose \vampire proofs cannot yet be checked by \lean, the reasons stem from large \lean files (column~4) or hard to reconstruct inferences and lacking support of some specific \vampire inference, implementation bugs or running into \lean tactic timeout (column~5).

 \cref{fig:results}\textbf{(a)} indicates that proof replay has a non-negligible influence on proofs that have been in generated in less time (shorter proving process). Rerunning few proof inferences for a shorter proof search is comparatively more costly when less inferences are made during proof search. Proof replay may take long time when there are lots of possible inferences from two premises. \cref{fig:results}\textbf{(b)}  shows that there is only minimal correlation between \vampire proof search and \lean proof reconstruction and checking. Finally, \cref{fig:results}\textbf{(c)} hints that longer proofs yield longer processes for proof reconstruction and checking times in \lean as one would expect. %
This is likely not only because the amount of theorem increase but also because longer proofs yield longer formulas to be checked via \vampire-specific \lean tactics that are not necessarily needed/optimized. %

\begin{figure}[H]
\begin{minipage}{.39\textwidth}
\begin{center}
\hspace{7mm}\textbf{(a)}
\includegraphics[height=5cm]{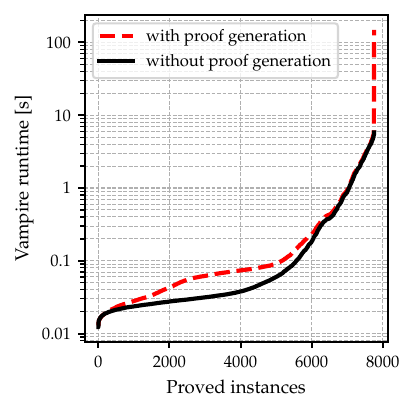}
\end{center}
\end{minipage}
\begin{minipage}{.3\textwidth}
\begin{center}
\hspace{7mm}\textbf{(b)}
\includegraphics[height=5cm]{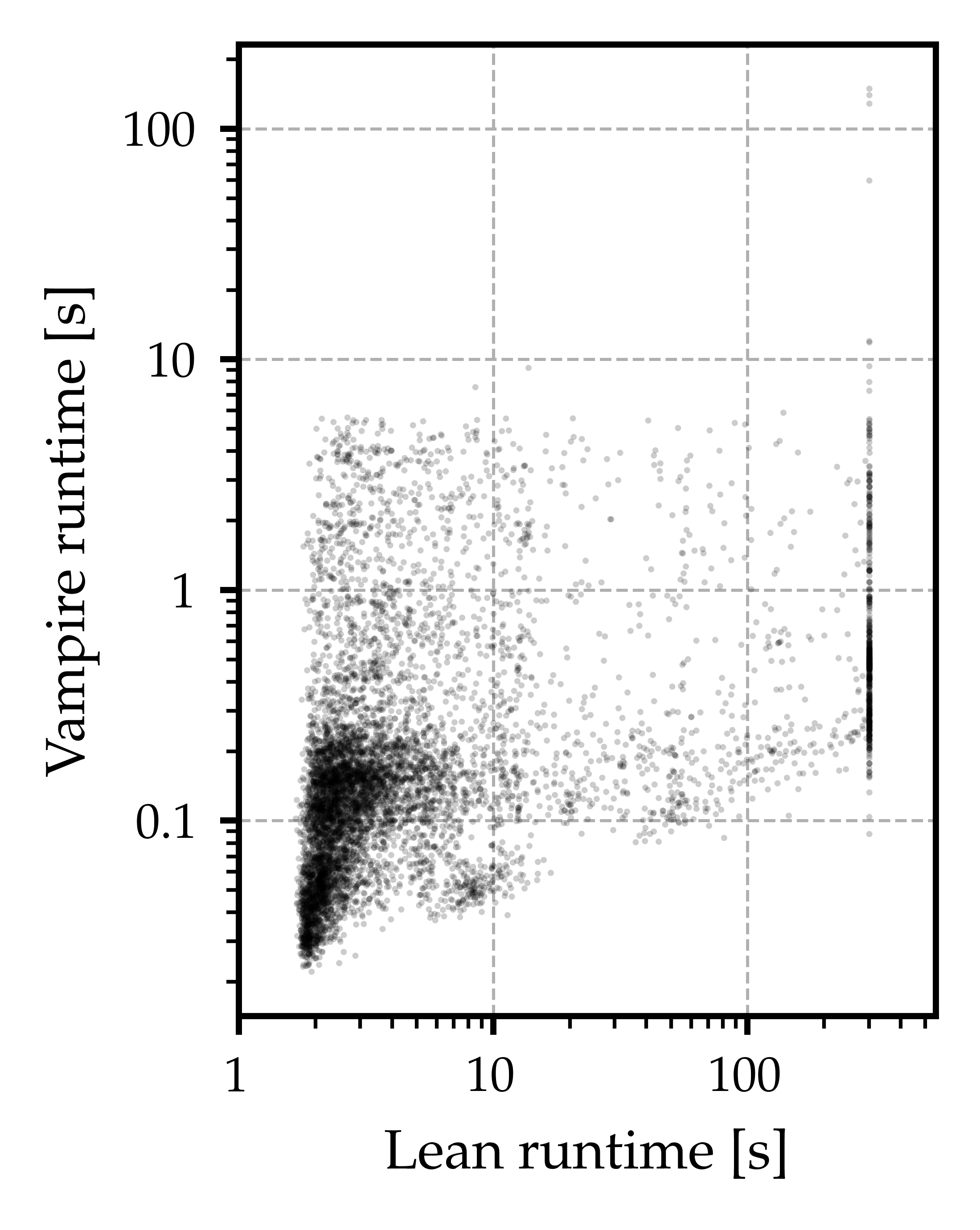}
\end{center}
\end{minipage}
\begin{minipage}{.3\textwidth}
\begin{center}
\hspace{7mm}\textbf{(c)}
\includegraphics[height=5cm]{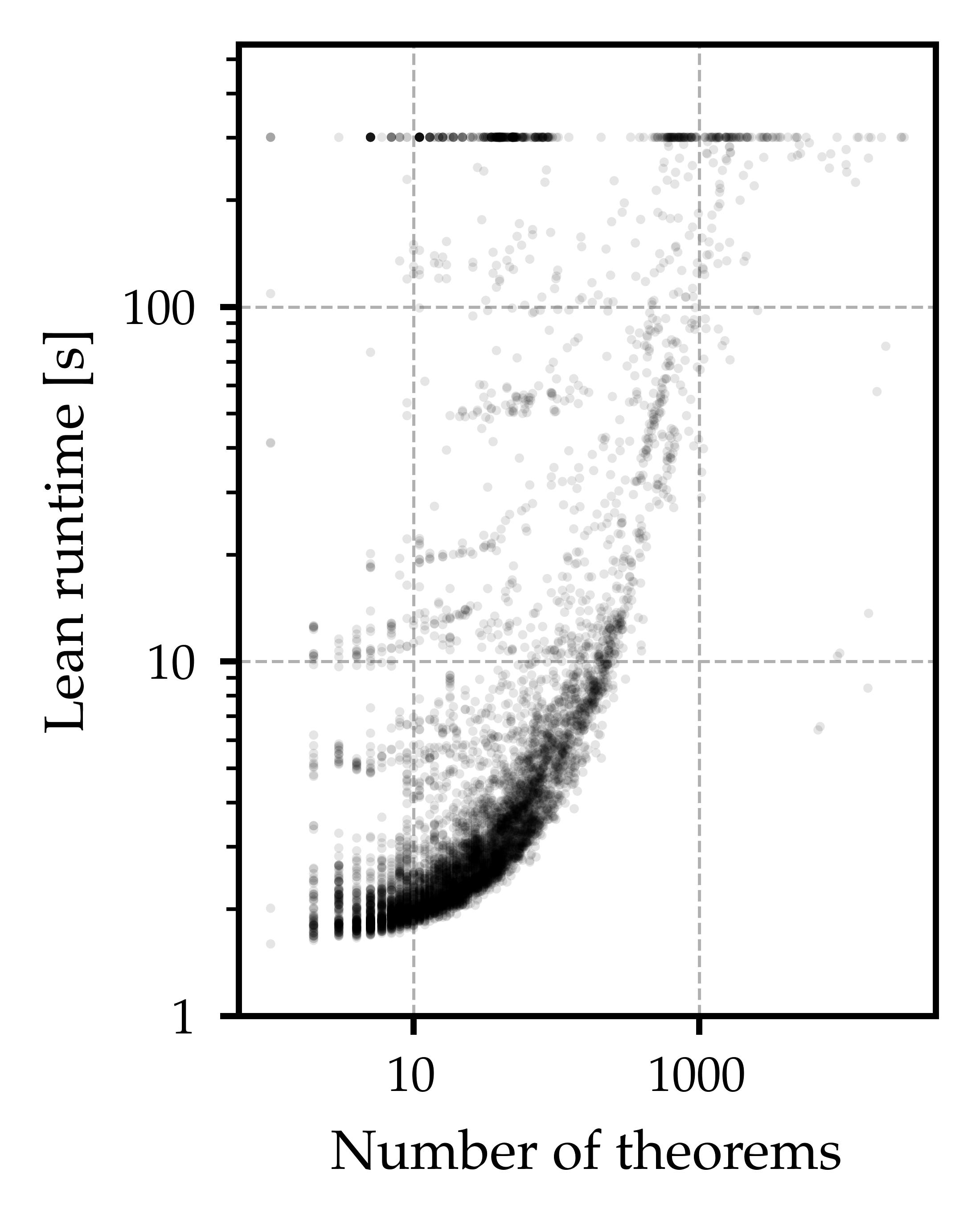}
\end{center}
\end{minipage}

\caption{\textbf{(a)} Comparing baseline \vampire and our \vampire version outputting a \lean file; here, we include  only CNF and FOF problems where baseline \vampire found a proof by contradiction. This plot only accounts for the overhead in \vampire, not for \lean proof reconstruction and checking time. Notice the logarithmic y-axis.
\textbf{(b)} A scatter plot comparing the \lean proof reconstruction and checking runtime vs the \vampire runtime. \textbf{(c)} A scatter plot of \lean runtime vs number of theorems in a \lean{} input file.
}
\label{fig:results}
\end{figure}

\end{section}

\section{Conclusions  %
and Future Work}
We describe ongoing efforts on reconstructing \vampire proofs in \lean, allowing us to trust  \vampire proofs  thanks to \lean's kernel (alongside a trusted SAT proof-checker \cite{bvdecide_tactic}). Our experiments   show  scalability of generating trusted \vampire proofs. 
We so far handle  only a small fraction of the inference steps implemented by \vampire. To remedy this, the proof-producing Duper ATP \cite{clune_et_al:LIPIcs.ITP.2024.10} is invoked if no specialized set of tactics is available. Further work plans addressing various  efficiency limits of proof reconstruction in \lean.  Notably,   %
the current implementation of CNF transformation  requires parsing and analyzing the same (possibly  large) formula multiple times. Saturation steps would also benefit by further information available within \vampire; these steps could be  utilized through specialized tactics in \lean.

\section{Use of Generative AI Tools}
Generative AI tools 
were not used in the writing of this paper. GitHub Copilot suggestions (that mainly rely on GPT5- mini) in VS Code have been used in our implementation. 

\bibliography{refs}

\end{document}